\documentstyle[12pt,a4wide]{article}

\begin{document}

\title{\bf Active galaxies can make axionic dark energy}

\author{Konstantinos Dimopoulos$^a$ and Sam Cormack$^b$\\
\\
$^a${\small\em Consortium for Fundamental Physics, Physics Department,}\\
{\small\em Lancaster University, Lancaster LA1 4YB, UK}
\\
\vspace{-.3cm}
\\
$^b${\small\em Department of Physics and Astronomy, Dartmouth College Hanover, 
NH 03755, USA}
\\
\vspace{-.3cm}
\\
{\small e-mails: {\tt k.dimopoulos1@lancaster.ac.uk}, \
{\tt samuel.c.cormack.gr@dartmouth.edu}}}

\maketitle

\newcommand{\lsim}{\mbox{\raisebox{-.9ex}{~$\stackrel{\mbox{$<$}}{\sim}$~}}}
\newcommand{\gsim}{\mbox{\raisebox{-.9ex}{~$\stackrel{\mbox{$>$}}{\sim}$~}}}

\begin{abstract}
AGN jets carry helical magnetic fields, which can affect dark matter if the 
latter is axionic. This preliminary study shows that, in the presence of strong
helical magnetic fields, the nature of the axionic condensate may change and 
become dark energy. Such dark energy may affect galaxy formation and galactic 
dynamics, so this possibility should not be ignored when considering axionic 
dark matter.
\end{abstract}

\section{Introduction}

As supersymmetric particles have not been observed in the LHC yet, interest in
axionic dark matter is increasing. Such dark matter has a loop-suppressed 
interaction with the electromagnetic field, which opens up observational 
possibilities that aim to exploit the photon-axion conversion in astrophysical 
magnetic fields. Many authors have considered the electromagnetic interaction 
of axion particles \cite{axion-photon}. However, the effect of this 
interaction to the axionic condensate itself has been largely ignored, assuming
that it is negligible. In this paper we investigate the effect of an
helical magnetic field on an axionic condensate. We find that, if the magnetic 
field is strong enough, axionic dark matter is modified to lead to the 
violation of the strong energy condition and behave as dark energy.%
\footnote{Axionic dark energy has been proposed before, see for example 
Ref.~\cite{alex}.}
Then we apply our findings to the helical magnetic fields in the jets of Active
Galactic Nuclei (AGN). We find that the magnetic fields near the central 
supermassive black hole may be strong enough to make axionic dark energy and 
thereby affect galaxy formation and dynamics. 
We use natural units, for which \mbox{$c=\hbar=1$} and \mbox{$m_P^{-2}=8\pi G$}, 
with \mbox{$m_P=2.4\times 10^{18}\,$GeV} being the reduced Planck mass. For the 
signature of the metric we take \mbox{$(+1,-1,-1,-1)$}.

\section{%
Electromagnetically dominated axionic condensate}

The axion (or an axion like particle) field $\phi$, at tree level has a 
coupling to a fermionic field $\psi$ of the form $\phi\bar\psi\psi$. Therefore,
the axion couples to the photon via a fermionic loop~as
\begin{equation}
{\cal L}_{\phi\gamma}=-\frac14 g_{\phi\gamma}\phi 
F_{\mu\nu}\tilde F^{\mu\nu}=g_{\phi\gamma}\phi \,\mbox{\boldmath $E$}\cdot
\mbox{\boldmath $B$}\,,
\label{LfA0}
\end{equation}
where $F_{\mu\nu}$ is the Faraday tensor, $\tilde F^{\mu\nu}$ is its dual and 
{\boldmath $E$} and {\boldmath $B$} are the electric and magnetic field 
respectively. In the above we have defined the dimensionful coupling 
\mbox{$g_{\phi\gamma}=\alpha\,{\cal N}/2\pi f_a$},
%\begin{equation}
%g_{\phi\gamma}=\frac{\alpha\,{\cal N}}{2\pi f_a}
%\end{equation}
where the $f_a$ is the Peccei-Quinn (PQ) scale, while 
\mbox{$\alpha\simeq 1/137$}. %In the simplest cases $\cal N$ is an integer, and 
We will assume ${\cal N}=1$. The Lagrangian density for the axion %and photon 
is
\begin{equation}
{\cal L}=\frac12\partial_\mu\phi\partial^\mu\phi-%\frac14F_{\mu\nu}F^{\mu\nu}-
%\frac12 m^2\phi^2
V(\phi)-\frac14 g_{\phi\gamma}\phi F_{\mu\nu}\tilde F^{\mu\nu},
\label{L}
\end{equation}
with
\begin{equation}
V(\phi)=m^2 f_a^2\left[1-\cos\left(\frac{\phi}{f_a}\right)\right],
\label{V}
\end{equation}
where $m$ is the axion mass.
\footnote{Such a potential has been utilised to realise cosmic inflation, 
in Natural Inflation originally in Ref.~\cite{natural} considering string 
axions. Much more recently, inflationary models considering also the axial, 
Chern-Simmons term $\propto F\tilde F$, have been considered in Gauge-flation 
\cite{Gauge-flation} and Chromo-natural Inflation \cite{ChromoNatural}. The 
axial term has also been employed to generate steep inflation with
parity-violating gravitational waves \cite{sorbo}.}
If inflation occurred after the PQ transition, the axion field is homogenised.  
\footnote{Alternatively, we may have production of a network of cosmic strings,
which however, we will not consider further since the scaling solution results 
in a few open strings per horizon volume anyway,
%.}
%\footnote{
so the field is aligned for
distances $\sim\,100\,$Mpc, unless one is near a cosmic string.} 
For a homogeneous axion in an expanding Universe, the Klein-Gordon equation of 
motion is
\begin{equation}
\ddot\phi+3H\dot\phi+
V'(\phi)
=g_{\phi\gamma}\,\mbox{\boldmath $E$}\cdot
\mbox{\boldmath $B$}\,,
\label{KG0}
\end{equation}
where $H(t)$ is the Hubble parameter, the dot denotes derivative with respect 
to the cosmic time and the prime denotes derivative with respect to $\phi$. 
In the early Universe the electromagnetic source term in the above may be 
neglected\footnote{unless an extremely strong, helical primordial magnetic 
field is present. We will not consider this possibility here.}.

Originally, the axion mass is zero and the vacuum manifold is flat. However,
at the quark confinement phase transition, QCD instantons tilt the vacuum 
manifold (for ${\cal N}=1$) and generate a sinusoidal potential for the
axion, as shown in Eq.~(\ref{V}). For sub-Planckian $f_a$, the axion begins 
coherently oscillating immediately after the phase transition, with original 
amplitude $\sim f_a$ \cite{KT}. However, very soon after the onset of the 
oscillations \mbox{$\phi\ll f_a$} and the potential becomes quadratic 
\mbox{$V\simeq\frac12 m^2\phi^2$}. Soon the period of oscillations 
becomes exponentially smaller than the Hubble time,  
so that the expansion of the Universe can be ignored for timescales shorter 
than $H^{-1}$. Then, Eq.~(\ref{KG0}) becomes
\mbox{$\ddot\phi+m^2\phi\simeq 0$} with solution
\begin{equation}
\phi=\Phi\cos(mt+\beta)\,,
\end{equation}
with $\Phi$ being the oscillation amplitude and $\beta$ being an initial phase.
Over cosmological timescales the Hubble friction term in Eq.~(\ref{KG0})
cannot be neglected, so that $\Phi=\Phi(t)$. A homogeneous scalar field 
oscillating coherently with a quadratic potential 
%corresponds to a collection
%of massive particles (axions) with zero momentum \cite{KT}, so that it 
can be regarded as pressureless matter whose density scales as \cite{KT}
\begin{equation}
\rho_\phi=\frac12m^2\Phi^2\propto a^{-3}\propto T^3
\Rightarrow
\Phi\sim f_a\left(\frac{T}{\Lambda_{\sc qcd}}\right)^{3/2},
\label{Phi}
\end{equation}
where 
%we have assumed that, initially, \mbox{$\Phi\sim f_a$}~, 
the scale factor is $a\propto 1/T$ and the temperature at the onset of the 
oscillations is determined by the scale of the QCD transition $\Lambda_{\sc qcd}$.
%\mbox{$T\sim\Lambda_{\sc qcd}$}~.
We have also assumed that $m$ remains constant but this is strictly speaking 
not true because \mbox{$m=m(T)$} initially grows while the oscillations are 
not harmonic near their onset, so Eq.~(\ref{Phi}) is off by a few orders of 
magnitude. In fact, the axion mass is estimated as \cite{KT}
\begin{equation}
m\sim\frac{\Lambda_{\sc qcd}^2}{f_a}\;\Rightarrow\;
%m\sim 10^{-5}
\frac{m}{10^{-5}\,{\rm eV}}\sim
%\left(
\frac{10^{12}\,{\rm GeV}}{f_a}\,,
%\right)%{\rm eV}
\label{mf}
\end{equation}
where \mbox{$\Lambda_{\sc qcd}\sim 100\,$MeV}.
Since the axion is to be the observed dark matter, its present density is
{$\rho_{\rm DM0}\simeq 0.3\rho_0\sim 10^{-30}\,$g/cm$^3$}. Because
\mbox{$\rho_{\rm DM0}=\frac12(m\Phi_0)^2$}, we find 
\mbox{$\Phi_0\sim 10^{-24}\,$GeV$^2/m$}, where $\Phi_0$ 
is the average value of $\Phi$ today. Using, Eq.~(\ref{mf}) we readily find
\begin{equation}
\Phi_0\sim 10^{-1}%\,{\rm eV}
%\frac{\Phi_0}{10^{-5}{\rm eV}}
\left(\frac{f_a}{10^{12}\,{\rm GeV}}\right){\rm eV}.
\label{Phif}
\end{equation}

Now, suppose that the electromagnetic source term in Eq.~(\ref{KG0}) cannot be
neglected. In this case, the oscillating axion may have a spatial dependence
as well, due to %the spatial dependence of 
the helical magnetic field. 
Considering timescales much smaller than $H^{-1}$ %the Hubble time 
and assuming Minkowski
spacetime for the moment, the equation of motion becomes
\begin{equation}
\ddot\phi-\nabla^2\phi+m^2\phi=
g_{\phi\gamma}\,\mbox{\boldmath $E$}\cdot
\mbox{\boldmath $B$}\,.
\label{KG}
\end{equation}
To keep things simple, let us consider that the electromagnetic source term has 
no time dependence. This implies that we consider the magnetic field of 
gravitationally bound structures (such as galaxies) which do not expand with 
the Universe expansion.
%(that is the helical magnetic field is roughly static, at least compared to 
%the period of oscillations $m^{-1}$). 
Then, the solution to the above is of the form
\begin{equation}
\phi=\Phi\cos(mt+\beta)+Q\,,
\label{phi}
\end{equation}
where $\dot Q=0$. Putting this in Eq.~(\ref{KG}) we have
\begin{equation}
(m^2-\nabla^2)Q=g_{\phi\gamma}\,\mbox{\boldmath $E$}\cdot\mbox{\boldmath $B$}\,.
\end{equation}
Now, let's assume that the characteristic
size of the helical magnetic field configuration is much larger than the
axion Compton wavelength. %$m^{-1}$. 
Then \mbox{$\nabla^2Q\ll m^2Q$}, so we find
\begin{equation}
Q\simeq g_{\phi\gamma}\,\frac{\mbox{\boldmath $E$}\cdot\mbox{\boldmath $B$}}{m^2}
\,.
\label{Q}
\end{equation}

For the energy-momentum tensor we have
\begin{equation}
T_{\mu\nu}=2\frac{\partial{\cal L}}{\partial g^{\mu\nu}}-g_{\mu\nu}{\cal L}
\label{TmnL}
\end{equation}
%Considering only the axion terms in the above, 
The axion-photon coupling in Eq.~(\ref{LfA0}) can be written as
\begin{equation}
{\cal L}_{\phi\gamma}=\frac12\,g_{\phi\gamma}\phi\;
\frac{\epsilon^{\mu\nu\rho\sigma}}{\sqrt{-g}}
F_{\mu\nu}F_{\rho\sigma}
%\tilde F^{\mu\nu}=g_{\phi\gamma}\phi \,\mbox{\boldmath $E$}\cdot
%\mbox{\boldmath $B$}\,,
\label{LfA}
\end{equation}
where $g\equiv\det(g_{\mu\nu})$ and $\epsilon^{\mu\nu\rho\sigma}$ is the Levi-Civita
symbol in flat spacetime ($=0,\pm 1$). Because 
\mbox{$\frac{\partial}{\partial g^{\mu\nu}}(\frac{1}{\sqrt{-g}})=\frac12
%g_{\mu\nu}/\sqrt{-g}
\frac{g_{\mu\nu}}{\sqrt{-g}}
$} it can be shown that 
\mbox{$2\frac{\partial{\cal L}_{\phi\gamma}}{\partial g^{\mu\nu}}=
g_{\mu\nu}{\cal L}_{\phi\gamma}$}. In view of Eq.~(\ref{TmnL}), this implies that 
${\cal L}_{\phi\gamma}$ does not contribute to $T_{\mu\nu}$.\footnote{This is why
${\cal L}_{\phi\gamma}$ is sometimes called ``topological''.} Na\"{i}vely, one
might think that the electromagnetic field cannot influence the energy-momentum
of the condensate, but this is incorrect because the effect of the
helical magnetic field is included in the $Q$-dependence of $\phi$ in 
Eq.~(\ref{phi}). 

Considering $\cal L$ as given in Eq~(\ref{L}) but ignoring 
${\cal L}_{\phi\gamma}$, we find for the energy density of the axion condensate
\begin{equation}
\rho_\phi=T_{00}=\frac12[\dot\phi^2+(\nabla\phi)^2]+\frac12 m^2\phi^2,
\label{rhophi}
\end{equation}
The pressure corresponds to the spatial components of $T_{\mu\nu}$.
The off diagonal spatial components are simply 
\mbox{$T_{ij}=\partial_i\phi\,\partial_j\phi$}, where \mbox{$i,j=1,2,3$} and
\mbox{$i\neq j$}. The diagonal components (the principal pressures) are
given by
\mbox{$T_{ii}=(\partial_i\phi)^2+p_\phi$} %(no Einstein summation is assumed) 
with
\begin{equation}
p_\phi={\cal L}-{\cal L}_{\phi\gamma}=
\frac12[\dot\phi^2-(\nabla\phi)^2]-\frac12 m^2\phi^2.
\label{pphi}
\end{equation}
The spatial gradients of the axion configuration mirror the spatial gradients 
of the magnetic field. If the characteristic dimensions of the latter are much 
larger than the axion Compton wavelength %$m^{-1}$ 
then, on average, 
\mbox{$|\partial_i\phi|\ll m|\phi|$}. 
This means that the off diagonal components of $T_{\mu\nu}$ become negligible and
the principal pressures all become approximately equal to $p_\phi$ in 
Eq.~(\ref{pphi}). We insert Eq.~(\ref{phi}) in the above to find
\begin{equation}
\bar\rho_\phi\simeq\frac12 m^2\Phi^2%+\frac12(\nabla Q)^2
+\frac12m^2Q^2\qquad{\rm and}\qquad
\bar p_\phi\simeq%-\frac12(\nabla Q)^2+
-\frac12m^2Q^2,
\label{rho}
\end{equation}
where $\bar\rho_\phi$ and $\bar p_\phi$ is the average density and pressure over
many oscillations such that \mbox{$\overline{\cos\omega}=0$}
and \mbox{$\overline{\sin^2\omega}=\overline{\cos^2\omega}=\frac12$} with 
\mbox{$\omega=mt+\beta$} and
we considered a magnetic field configuration with typical 
scale larger than %the Compton wavelength of the condensate 
$m^{-1}$ so that \mbox{$(\nabla Q)^2\ll (m\,Q)^2$}.
%and also \mbox{$|\partial_\parallel Q|,|\partial_\perp Q|\ll m\,Q$}.

It is evident that, in the absence of a magnetic field, we have 
\mbox{$\bar p_\phi=0$} so the axionic condensate behaves as dark matter with 
\mbox{$\rho_\phi=\frac12 m^2\Phi^2$}. However, we will assume that a helical
magnetic field is present. If the magnetic field contribution in the above is
dominant (i.e. \mbox{$\Phi^2<2Q^2$}) we have 
\mbox{$\bar p_\phi/\bar\rho_\phi<-\frac13$}
and the axionic condensate violates the strong energy condition and behaves as 
dark energy. In effect, when \mbox{$Q^2\gg\Phi^2$} the helical magnetic field 
halts in its track the oscillating condensate such that \mbox{$\phi\simeq Q$}. 
The condensate density becomes 
\mbox{$\rho_\phi\simeq V(Q)\simeq\frac12m^2Q^2\simeq\,$constant} (in time).
Thus, the axionic condensate becomes potentially dominated and acts similarly to
the inflaton condensate during slow-roll inflation.
Note however that, from Eq.~(\ref{rho}) we have
\mbox{$\bar\rho_\phi+\bar p_\phi=\frac12m^2\Phi^2>0$}, as with the case of 
axionic dark matter since the contribution of the magnetic field cancels out.%
\footnote{This can be readily seen from Eqs.~(\ref{rhophi}) and (\ref{pphi}), 
which give \mbox{$\rho_\phi+p_\phi=\dot\phi^2=m^2\Phi^2\sin^2\omega$}.}
Thus, the condensate does not violate the null energy condition. 

A non-oscillatory scalar filed condensate does not have a particle 
interpretation. So, what happens to the axion particles, which the oscillatory 
condensate coresponds to, when an intense helical magnetic field is present? 
Eq.~(\ref{phi}) suggests that, even when $Q$ is sizable, the axion oscillations
continue, with the same frequency but not arround zero; around $Q$ instead.
Because the effect of the helical magnetic field is additive in Eq.~(\ref{phi}),
we expect that the axion dark matter particles coexist with a smooth axion field
component, like photons travelling inside a constant electromagnetic field.
The smooth axionic condensate is gravitationally repulsive (as dark energy is)
so we would expect axion dark matter particles to be driven away from the 
axionic dark energy.

Therefore, the conditions necessary for turning 
axionic dark matter into dark energy are
\begin{equation}
(\nabla Q)^2\ll (m\,Q)^2\qquad{\rm and}\qquad 2Q^2>\Phi^2,
\label{conditions}
\end{equation}
where $Q$ is determined by the helical magnetic field as shown in Eq.~(\ref{Q}).

\section{Axionic matter in AGNs}

It so happens that the conditions in Eq.~(\ref{conditions}) may be satisfied in
AGN jets. Observations suggest that AGN jets feature powerful helical 
magnetic fields \cite{helical}. 
Most spiral galaxies are assumed to go through the AGN phase 
when their central supermassive black hole is formed. 
%\footnote{In fact, spiral galaxies may be the remnants of the accretion disk 
%from the AGN phase \cite{AGNbook}.} 
The AGN jet can be huge in length (up to Mpc scales). Its spine, however, is 
narrow; about \mbox{$d\sim(10^{-5}-50)\,$pc}, which, however, is typically 
much larger than the axion Compton wavelength $m^{-1}$.
Therefore, $(\nabla Q)^2\ll (m\,Q)^2$ is satisfied. Now, the AGN spine is 
electromagnetically dominated, so that $2Q^2>\Phi^2$ seems reasonable. 
Let us estimate~this.

The helicity of the magnetic field along the AGN jet is thought to be due to 
the rotation of the accretion disk \cite{AGNbook}. It can be modelled as a 
longitudinal (poloidal) field {\boldmath $B_\parallel$} and a transverse 
(toroidal) field {\boldmath $B_\perp$}, which is a Biot-Savart field due to the 
current along the jet, such that 
\mbox{{\boldmath $J$}$\,=\nabla\times\,${\boldmath $B_\perp$}}, where
\mbox{{\boldmath $J$}$\,=\sigma\,${\boldmath $E$}} is the current density and 
$\sigma$ is the plasma conductivity. Thus, from Eq.~(\ref{Q}) we have 
\begin{equation}
Q=\frac{\alpha\eta}{2\pi f_am^2}\,
(\nabla\times\mbox{\boldmath $B_{\perp}$})
\cdot\mbox{\boldmath $B$}\;\Rightarrow\;|Q|\sim
\frac{\alpha\eta}{f_am^2}\frac{B_\perp\,B_\parallel}{d},
\label{Q0}
\end{equation}
where $\eta=1/\sigma$ is the plasma resistivity. The above needs to be 
compared with $\Phi$, the amplitude of the oscillating axionic condensate.

The average amplitude at present is given by Eq.~(\ref{Phif}).
However, inside galaxies, the dark matter density is expected to be much larger
than the average density of dark matter at present $\rho_{\rm DM0}$.
Indeed, a typical estimate for dark matter in the vicinity of the Earth is
\mbox{$\rho_{\rm DME}\simeq 0.3\,$GeV/cm$^3$}, which is about $10^5\rho_{\rm DM0}$.
Depending on the dark matter halo model, the density of dark matter can be 
$\sim$~50 times bigger in the core.
%, and if a central cusp is assumed, this may be even larger. 
Because \mbox{$\rho_{\rm DM}\propto\Phi^2$} this means that the 
core value of $\Phi$ can be \mbox{$\Phi_{\rm C}\sim 10^3\Phi_0$}.

From Eqs.~(\ref{mf}) , (\ref{Phif}) and (\ref{Q0}), it is straightforward to 
show that
\begin{equation}
\left.\frac{|Q|}{\Phi}\right|_{\rm C}\sim
\frac{\eta}{d}\frac{(B_\perp B_\parallel)_{\rm core}}{10^{14}\,{\rm G}^2}\,,
\end{equation}
where the dependence on the Peccei-Quinn scale $f_a$ cancels out.

The onset of the jet is near the event horizon. To suppress relativistic 
corrections, we choose \mbox{$r_0\sim 10^2\,r_S$} for the onset of the
jet, where \mbox{$r_S=2GM_{\rm BH}$} is the Scwarzschild radius 
\begin{equation}
r_S\sim 10^{-5}\left(\frac{M_{\rm BH}}{10^8M_\odot}\right){\rm pc}\,.
\label{rS}
\end{equation}

Now, flux conservation suggests that \mbox{$B_\perp\propto 1/r$}
and \mbox{$B_\parallel\propto 1/r^2$}, where $r$ %runs along the jet spine and 
denotes the distance from the central supermassive black hole 
\cite{AGNbook,AGNpaper}. 
Therefore,
\begin{equation}
\frac{(B_\perp B_\parallel)_{\rm core}}{(B_\perp B_\parallel)_{\rm lobe}}
\sim\left[\frac{\rm 1\,Mpc}{\rm 10^{-3}\,pc}
\left(\frac{10^8M_\odot}{M_{\rm BH}}\right)\right]^3
= 10^{27}\left(\frac{10^8M_\odot}{M_{\rm BH}}\right)^3,
\label{Bs}
\end{equation}
where we assumed that the jet is about 1~Mpc long. The {\em minimum} magnetic 
field at the lobes is of equipartition value $\sim\mu$G \cite{AGNbook}. So, 
\mbox{$(B_\perp B_\parallel)_{\rm lobe}\sim 10^{-12}\,$G$^2$}. 
%Note that 
%\mbox{$(B_\perp B_\parallel)_{\rm core}\sim 
%10^{18}(B_\perp B_\parallel)_{\rm lobe}\sim 
%10^9\,$G$^2$} nicely agrees with the fact that magnetic fields of about 
%$10^4\,$G are expected in the accretion disk. 
Therefore, 
\begin{equation}
\left.\frac{|Q|}{\Phi}\right|_{\rm C}\sim 10\,(\eta/d)\times 
\left(\frac{10^8M_\odot}{M_{\rm BH}}\right)^3.
\label{QPhi}
\end{equation}

The resistivity %(= diffusivity in natural units) 
of the plasma is still an open
issue, although it is implicitly estimated in MHD simulations. This is because
the plasma in jets is far from thermal equilibrium, while it is still uncertain
which particles are the primary charge carriers for the current. An estimate, 
based on numerical investigation, is provided by \cite{tsinganos}
\begin{equation}
\frac{\eta}{10^{14}\,{\rm m^2s^{-1}}}\sim 
\sqrt{\frac{M_{\rm BH}}{M_\odot}\frac{r_0}{r_\odot}}
\;\Rightarrow\;
\eta\sim 10^{-4}\,{\rm pc}
\times\left(\frac{M_{\rm BH}}{10^8M_\odot}\right),
\end{equation}
where we considered $r_0\sim 10^2\,r_S$.
%Also, the conductivity of the plasma is 
%\mbox{$\sigma\sim 10^{-10}\,m_e/e^2$}, where $m_e$ and $e$
%%\mbox{$m_e\sim 0.5\,$MeV} and \mbox{$e\simeq 0.3$} 
%are the electron mass and electric charge. %respectively.
%Then, the plasma resistivity is 
%$\eta%=1/\sigma%\sim 10^{10}\,$MeV$\,
%\sim 10^{-19}\,$pc. 
Assuming the width of the jet spine near the core is 
\mbox{$d\sim r_0$}, we have
%\mbox{$\eta/d\sim 10^{-15}$}. %In this case, Eq.~(\ref{QPhi}) becomes
%Putting everything together we find
\begin{equation}
\left.\frac{|Q|}{\Phi}\right|_{\rm C}\sim 
\left(\frac{10^8M_\odot}{M_{\rm BH}}\right)^3.
\label{QPhiFinal}
\end{equation}
Thus, we see that the axionic condensate can become dark energy near the 
AGN core if \mbox{$M_{\rm BH}\lsim 10^8M_\odot$}, which is quite plausible.
For example, the supermassive black hole in the centre of the Milky Way 
has mass $\sim 10^6M_\odot$.% 
\footnote{Surprisingly, the 
effect is intensified for smaller black holes because the horizon size is 
smaller, while the AGN jets are always taken to be of Mpc scales. This, might 
be augmented when more realistic AGN jets are considered.}

Furthermore, from Eq.~(\ref{mf}) we find 
\begin{equation}
m^{-1}\sim 10^{-18}\left(\frac{f_a}{10^{12}\,{\rm GeV}}\right){\rm pc}
\end{equation}
Therefore, \mbox{$m^{-1}\ll d$} for sub-Planckian $f_a$ and the magnetic field 
configuration is safely much larger than the axionic Compton wavelength.

In general, we expect to be in Kerr spacetime. Jets are projected 
along the directions of the poles, which implies 
\mbox{$\sin\theta\simeq 0$}. Then, the Kerr metric reduces to
\begin{equation}
ds^2=\frac{\Delta}{r^2}dt^2-\frac{r^2}{\Delta}dr^2-r^2d\theta^2,
\end{equation}
where \mbox{$\frac{\Delta}{r^2}=1-\frac{r_S}{r}+\left(\frac{a}{r}\right)^2$}, 
with \mbox{$a=GJ/M_{\rm BH}$}; $J$ being the angular momentum. Assuming that 
the black hole is not extremal means \mbox{$a<r_S$}. So, taking
\mbox{$d\sim r_0\sim 10^2\,r_S$} ensures that relativistic corrections are 
small.

\section{Conclusions}

We have investigated the behaviour of axionic dark matter in the presence of
a helical magnetic field and found that, when the condensate becomes
electromagnetically dominated, it ceases to be dark matter and becomes dark 
energy instead. We have applied our findings in AGNs and showed that the 
helical magnetic field along the AGN jets near the AGN core can be strong 
enough to convert axionic dark matter into dark energy. Lacing the AGN black 
holes with dark energy may have profound implications for galaxy formation 
and galactic dynamics (e.g. rotation curves), because dark energy is 
gravitationally repulsive. 
%\footnote{It is an open question whether gravitational repulsion 
%can drive away the axionic condensate from the supermassive black hole.} 
Note that, in contrast to the evenly distributed dark energy, which dominates 
the Universe today, the axionic dark energy in AGNs is localised in the core
parts of the AGN jets. Our study 
%is not thorough enough to provide definite answers but 
demonstrates that the issue warrants deeper investigation.
It is likely that relativistic corrections may influence the results.

\subsubsection*{%
%\medskip
%
%\noindent
%{\bf 
Acknowledgements}

\noindent
KD would like to thank A.C.~Fabian, I.M.~Hook, D.H.~Lyth, J.~McDonald and
T.~Raptis for discussions and comments. This work was supported (in part) by the
Lancaster-Manchester-Sheffield 
Consortium for Fundamental Physics under STFC grant ST/L000520/1.

\end{document}